\documentclass[prd,nofootinbib]{revtex4}
\usepackage{amsmath}
\usepackage{graphics}
\usepackage{epsfig}
\begin{document}
\title{Process $\eta \to \pi^0 \gamma \gamma$ in the Nambu--Jona-Lasinio model}

\author{A. E. Radzhabov}%
\email{aradzh@theor.jinr.ru}
\author{M. K. Volkov}
\email{volkov@theor.jinr.ru}

\affiliation{%
Bogoliubov Laboratory of Theoretical Physics, Joint Institute for Nuclear Research,
141980 Dubna, Russia }

\date{\today}

\begin{abstract}
The decay width of the process $\eta \to \pi^0 \gamma \gamma$ is calculated in the
framework of the Nambu--Jona-Lasinio model. The momentum dependence of quark loops
is taken into account. Three types of diagrams are considered: quark box,
scalar($a_0$) and vector($\rho,\omega$) pole diagrams. The obtained estimations are
in satisfactory agreement with recent experimental data.

\end{abstract}
\maketitle

The investigations of the process $\eta \to \pi^0 \gamma \gamma$ have a long
history. The experimental studies of this process began in 1966\footnote{Excellent
review of theoretical and experimental works can be found in \cite{Achasov:2001qm}.}
\cite{PhysRevLett.16.767}. The first experimental results led to a large value of
the branching ratio of the process. The theoretical estimates obtained in the vector
dominance model (VDM) \cite{PhysRev.160.1397}, nonlinear chiral
theory\footnote{Note that similar result for the width of the order $10^{-2}$ eV lately
obtained in ChPT for pion-loop contribution at the level $O(p^4)$.} \cite{Ebert:1979fe}
and lately in the chiral quark model
\cite{Ivanov:1982mz,Kreopalov:1982js,Volkov:1986zb} predicted noticeably lower
value.

A real breakthrough in the investigation of this process happened in the experiment GAMS in
1981 at Protvino \cite{Binon:1981xq} where the large energies of the produced $\eta$-mesons
dramatically suppressed the background. During a subsequent reanalysis, the value
$\Gamma_{\eta \to \pi \gamma \gamma}=0.84 \pm 0.18$ eV was obtained\footnote{Notice that this
result is consistent with those obtained in the NJL model
\cite{Kreopalov:1982js}.}\cite{Alde:1984wj}. Lately, the SND collaboration in the experiment
VEPP-2M confirmed this value to be an upper limit 1 eV for the width of the process
\cite{Achasov:2001qm}. In 2005, the results obtained by the Crystal Ball collaboration at BNL
AGS were published; these results $\Gamma_{\eta \to \pi \gamma \gamma}=0.45 \pm 0.12$
\cite{Prakhov:2005vx} were noticeably smaller than those reported by the GAMS collaboration.

From a theoretical point of view this process was investigated in many theoretical models: VDM
model \cite{PhysRev.160.1397}, nonlinear chiral theory \cite{Ebert:1979fe}, different quark
models \cite{Ivanov:1982mz,Kreopalov:1982js,Volkov:1986zb,Ng:1993sc,Nemoto:1996bh}, resonance
exchange models \cite{Ng:1992yg,Ko:1992zr}, the chiral perturbation theory (ChPT)
\cite{Ametller:1991dp,Ko:1993rg,Bellucci:1995ay,Bel'kov:1995fj,Bijnens:1995vg}, and chiral
unitary approach \cite{Oset:2002sh}. In ChPT, the main contribution comes from the terms of
the order O($p^6$) of low energy expansion because the tree terms of the order O($p^2$) and
O($p^4$) are absent and one-loop contributions of the order O($p^4$) are very small. The
counterterms of the order O($p^6$) are not determined from the theory itself and should be
fixed using experimental information, from the assumption of meson saturation (vector meson
exchange giving the dominant contribution) or calculated from the model (NJL for example). In
\cite{Ametller:1991dp}, the meson saturation approach was adopted, which gave $\Gamma_{\eta
\to \pi \gamma \gamma}=0.18$ eV; too small, compared to the experimental value. But, keeping
the momentum dependence in the vector meson propagators gives an ``all-order'' estimate of
about 0.31 eV \cite{Ametller:1991dp}, in agreement with the old VDM prediction
\cite{PhysRev.160.1397}. Taking into account the scalar and tensor meson contributions (the
signs of which cannot be unambiguously determined within this approach) and the one-loop
contribution at O($p^8$), the final estimate of \cite{Ametller:1991dp} is $\Gamma_{\eta \to
\pi \gamma \gamma}= 0.42\pm 0.20$ eV, in a satisfactory agreement with the recent Crystal Ball
result. This result is confirmed in \cite{Bellucci:1995ay}, where the O($p^6$) counterterms
are calculated in the framework of the NJL model with the result $0.58\pm0.3$ eV. However, the
same counterterms obtained from the NJL model by different methods lead to $0.1$ eV
\cite{Bel'kov:1995fj} and $0.27^{+0.18}_{-0.07}$ eV \cite{Bijnens:1995vg}.

The ``all-order'' estimations in \cite{Ametller:1991dp} are a signal that the preservation of
full momentum dependence is highly desirable. Note that in
\cite{Kreopalov:1982js,Volkov:1986zb} the simple NJL model is used without taking into account
the momentum dependence of quark loops. Then, in a quark models
\cite{Ng:1993sc,Nemoto:1996bh}, the full momentum dependence of the quark box diagram is
considered whereas the diagram with the intermediate scalar $a_0$(980) is ignored. The vector
sector of the model has not been taken into account as well.

In the present work, the process $\eta\to\pi\gamma\gamma$ is calculated in the framework of
the NJL model with scalar--pseudoscalar and vector--axial-vector sectors. The contribution of
the quark box loop is considered together with the contributions of the diagrams with scalar
and vector intermediate mesons (as in \cite{Kreopalov:1982js,Volkov:1986zb}). The momentum
dependence of the quark loops and pseudoscalar--axial-vector transitions are taken into
account, following \cite{Bernard:1992mp,Bernard:1995hm,Bajc:1996gt}.

\section{The $U(3)\times U(3)$ NJL model}

The $U(3)\times U(3)$ NJL model with scalar-pseudoscalar and vector-axial-vector sectors is
used in the present work. To solve the $U_A(1)$ problem, the six-quark t`Hooft interaction is
added to the Lagrangian of the model \cite{Klimt:1989pm,Klevansky:1992qe}
\begin{eqnarray}
 \mathcal{L}& =& {\bar q}(i{\hat \partial} - m^0)q
  + \frac{G}{2}\sum_{i=0}^8 [({\bar q} {\lambda}_i q)^2 +({\bar q}i{\gamma}_5{\lambda}_i q)^2]
  + \frac{G_V}{2}\sum_{i=0}^8 [({\bar q} {\gamma}_\mu {\lambda}_i q)^2 +({\bar q}{\gamma}_5{\gamma}_\mu{\lambda}_i  q)^2]
   \nonumber \\
 &&- K \left( {\det}[{\bar q}(1+\gamma_5)q]+{\det}[{\bar q}(1-\gamma_5)q] \right),
\label{Ldet}
\end{eqnarray}
where $\lambda_i$ (i=1,...,8) are the Gell-Mann matrices and $\lambda^0 =
{\sqrt{\frac{2}{3}}}${\bf 1}, with {\bf 1} being the unit matrix; $m^0$ is the current quark
mass matrix with diagonal elements $m^0_u$, $m^0_d$, $m^0_s$ $(m^0_u \approx m^0_d)$, $G$ and
$G_V$ are the scalar--pseudoscalar and vector--axial-vector four-quark coupling constants; $K$
is the six-quark coupling constants. The six-quark interaction can be reduced to an effective
four-fermion vertex after the contraction of one of the quark pairs. The details are given in
appendix A.

Light current quarks transform to massive constituent quarks as a result of spontaneous chiral
symmetry breaking. Constituent quark masses can be found from the Dyson-Schwinger equation for
the quark propagators (gap equations)
\begin{eqnarray}
m_u&=&m_u^0 + 8 m_u G I_1(m_u)+32 m_u m_s K I_1(m_u) I_1(m_s)\nonumber\\
m_s&=&m_s^0 + 8 m_s G I_1(m_s)+32 K \left(m_uI_1(m_u)\right)^2, \label{gapNJL}
\end{eqnarray}
where $I_1(m)$ is the quadratically divergent integral. The modified Pauli-Villars (PV)
regularization with two substractions with same $\Lambda$ is used for the regularization of
divergent integrals\footnote{Any function $f(m^2)$ of mass $m^2$ is regularized by using the
rule
\begin{eqnarray}
f(m^2)\to f(m^2)-f(m^2+\Lambda^2)+\Lambda^2 f^\prime(m^2+\Lambda^2).\nonumber
\end{eqnarray}} (see
\cite{Bernard:1992mp,Bernard:1995hm,Bajc:1996gt,Schuren:1991sc}). In this case the
quadratically and logarithmically divergent integrals $I_1(m)$ and $I_2(m)$ have the same form
as in the four-momentum cut-off scheme
\begin{eqnarray}
I_1(m) &=& \frac{N_c}{4 \pi^2}
\left[\Lambda^2-m^2\ln\left(\frac{\Lambda^2}{m^2}+1\right)\right], \quad
I_2 (m) =\frac{N_c}{4 \pi^2}
\left[\ln\left(\frac{\Lambda^2}{m^2}+1\right)-\left(1+\frac{m^2}{\Lambda^2}\right)^{-1}\right]
\nonumber.
\end{eqnarray}

Moreover, the Pauli-Villars regularization is suitable for the description of the vector
sector because it preserves gauge invariance.

Masses and vertex functions of the mesons can be found from the Bethe-Salpeter equation. The
expression for the quark-antiquark scattering matrix is
\begin{eqnarray}
\hat{T}=\mathbf{G}+\mathbf{G}\mathbf{\Pi}(p^2)\hat{T}=\frac{1}{\mathbf{G}^{-1}-\mathbf{\Pi}(p^2)},
\end{eqnarray}
where $\mathbf{G}$ and $\mathbf{\Pi}(p^2)$ are the corresponding matrices of the four-quark
coupling constant and polarization loops. The particle mass can be found from the equation
$\mathrm{det}(\mathbf{G}^{-1}-\mathbf{\Pi}(M^2))=0$ and near the poles the corresponding part
of the $\hat{T}$ matrix can be expressed in the form
\begin{eqnarray}
\hat{T}=\frac{\bar{V} \otimes V}{p^2-M^2},
\end{eqnarray}
where $V$ and $M$ are the vertex function and mass of the meson, and $\bar{V} = \gamma^0
V^\dag \gamma^0$. Details of calculations for different channels are presented in appendices
B, C. Here we discuss only general properties.

The most simple situation takes place for the vector and the isovector scalar meson with equal
quark masses (say $\rho$ and $a_0$). In this case, the coupling constant and polarization
operator are just numbers (not matrices). For pseudoscalar mesons, additional axial-vector
components appear in the vertex function due to the pseudoscalar--axial-vector mixing (in the
scalar case this transition loop is proportional to the difference of quark masses). An
additional complication takes place for $\eta$ and $\eta^\prime$ due to the singlet-octet
mixing (or mixing of strange and non-strange quarks due to the t`Hooft interaction).
Therefore, the vertex function of this meson has four components: strange and non-strange
pseudoscalar and axial-vector.

\section{Fixing model parameters}

The model has six parameters: the coupling constants $G$, $G_V$, $K$, PV cut-off $\Lambda$,
and constituent quark masses $m_u$ and $m_s$. We use two parametrization schemes. In the first
one, the model parameters are defined using masses of the pion, kaon, $\rho$ and $\eta$ mesons
and the weak pion decay constant $f_\pi$. Note that the number of input parameters is greater
than the number of physical observables by one. This allows us, following
\cite{Bernard:1995hm}, to take the mass of the $u$ quark slightly larger than the half of the
$\rho$-meson mass.
% which exclude unphysical thresholds of quark pair creation from quark loops.
As a result, we have the following set (set I) of model parameters
\begin{eqnarray}
 m_u = 390\,\mathrm{MeV},\,
 m_s=496\,\mathrm{GeV},\,
 G=6.62\,\mathrm{GeV}^{-2},\,
 G_V=-11.29\,\mathrm{GeV}^{-2},\,
 K = 123\,\mathrm{GeV}^{-5},\,
\Lambda=1\,\mathrm{GeV}.
\end{eqnarray}
The values of the current quark masses $m^0_u,m^0_s$ are defined from the gap
equations (\ref{gapNJL}) $m^0_u=3.9$ MeV and $m^0_s=70$ ($m^0_u/m^0_s=18$).

For this set of model parameters, the two-photon decay width of the $\eta$ meson
$\Gamma_{\eta\to \gamma\gamma}=0.37$ KeV, is smaller than the experimental one:
$\Gamma^{\mathrm{exp}}_{\eta\to\gamma\gamma}=0.510\pm0.026$ \cite{PDBook}.

In the set II the model parameters are fixed in order to reproduce the two-photon decay width
of the $\eta$ meson instead of its mass (the $\eta$ meson mass in this case $M_\eta=530$ MeV)
\begin{eqnarray}
 m_u = 390\,\mathrm{MeV},\,
 m_s=506\,\mathrm{GeV},\,
 G=8.04\,\mathrm{GeV}^{-2},\,
 G_V=-11.29\,\mathrm{GeV}^{-2},\,
 K = 77\,\mathrm{GeV}^{-5},\,
\Lambda=1\,\mathrm{GeV}.
\end{eqnarray}
The current quark masses are $m^0_u=3.9$ MeV and $m^0_s=78$ MeV ($m^0_u/m^0_s=20$).

\section{Decay $\eta \to \pi^0 \gamma \gamma$}
\begin{figure}
\resizebox{0.8\textwidth}{!}{\includegraphics{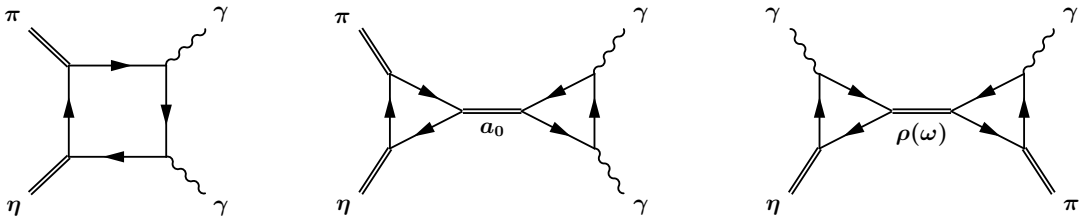}} \caption{\label{fig:etpigaga}
Diagrams contributing to the amplitude of the process $\eta \to \pi^0 \gamma \gamma$.}
\end{figure}

The general form of the $\eta \to \pi^0 \gamma \gamma$ decay amplitude contains two
independent tensor structures \cite{Ecker:1987hd}
\begin{eqnarray}
T= T^{\mu\nu}\epsilon^{1}_\mu\epsilon^{2}_\nu,\quad T^{\mu\nu}=A(x_{1},x_{2})(q_{1}^{\nu}
q_{2}^{\mu} - q_{1} \cdot q_{2} g^{\mu\nu}) + B(x_{1},x_{2}) \left[ -M_{\eta}^{2} x_{1}x_{2}
g^{\mu\nu} - \frac{q_{1} \cdot q_{2}}{M_{\eta}^{2}} p^{\mu}p^{\nu} + x_{1} q_{2}^{\mu} p^{\nu}
+ x_{2} p^{\mu} q_{1}^{\nu} \right], \label{ampform}
\end{eqnarray}
where $p$, $q_1$, $q_2$ are the momentum of the $\eta$ meson and photons, $\epsilon^{1}_\mu$
and $\epsilon^{2}_\nu$ are the polarization vectors of the photons, and $x_i= p\cdot
q_i/M_\eta^2$.

The $\eta \to \pi^0 \gamma \gamma$ decay width has the form
\begin{eqnarray}
 \Gamma & = & \frac{M_{\eta}^{5}}{256 \pi^{2}}
  \int\limits_{0}^{(1-y)/2} dx_{1} \int\limits_{x_2^{\mathrm{min}} }^{x_2^{\mathrm{max}}} dx_{2}
  \left\{ \left| A(x_{1},x_{2}) + \frac{1}{2} B(x_{1},x_{2}) \right|^{2} \left[ 2(x_{1}+x_{2})
+y -1 \right]^{2} \right. \nonumber \\
  &  + &  \left. \frac{1}{4} \left| B(x_{1},x_{2}) \right| ^{2} \left[ 4 x_{1} x_{2}
- \left[ 2(x_{1}+x_{2})+ y-1 \right] \right] ^{2} \right\}  ,\\
&&x_2^{\mathrm{min}}={(1-2x_1-y)/2 },\quad x_2^{\mathrm{max}}={(1-2x_1-y)/2(1-2x_1)},
\quad y =M_\pi^2/M_\eta^2.\nonumber
\end{eqnarray}

In the NJL model the amplitude for the $\eta \to \pi^0 \gamma \gamma$ decay process is
described by three types of diagrams (see Fig. \ref{fig:etpigaga}): the quark box and exchange
of scalar($a_0$) and vector ($\rho,\omega$) resonances. Let us consider theses contributions
in detail.

The scalar meson exchange has the simplest form. It gives a contribution only to $A(x_1,x_2)$.
This contribution consists of three parts and can be written in the form (see appendices B and
C for the definition of polarization loops and vertex functions):
\begin{eqnarray}
 A(x_1,x_2) &=& \frac{g_{a_0 \eta \pi}(2q_1\cdot q_2)g_{a_0 \gamma \gamma}(2q_1\cdot q_2)}{G_{a_0}^{-1}-\Pi_{SS}^{uu}(2 q_1\cdot q_2 )}
, \quad q_1\cdot q_2 = M_{\eta}^{2}\left(x_1+x_2-\frac{1}{2}\right)+\frac{M_\pi^2}{2}\nonumber\\
g_{a_0\gamma\gamma}(p^2)&=& \frac{1}{2 \pi^2}  \int \limits_0^1 dx_1 \int \limits_0^{1-x_1}
dx_2 \frac{m_u(1-4x_1x_2)}{(p^2 x_1 x_2-m_u^2-\Lambda^2)^2 (p^2 x_1 x_2-m_u^2)} \\
g_{a_0 \eta \pi}(p^2)&=& -i 2 N_c N_f\int \frac{d^4_\Lambda k}{(2\pi)^4}
  \mathrm {Tr}_D\left\{V_{a_0}S_u(k+q_1)V_{\pi}S_u(k)V_{\eta}S_u(k-q_2)\right\}.\nonumber
\end{eqnarray}
here $\mathrm {Tr}_D$ is the trace over Dirac indices, index $\Lambda$ in the measure of
integration means PV regularization of the integral and $S_j(p)=(\hat{p}-m_j)^{-1}$.

The amplitude with the vector meson ($\rho,\omega$) exchanges consists of two quark triangles
of anomalous type (see appendix D) and the vector meson propagator. It gives the following
contributions
\begin{eqnarray}
 B(x_{1},x_{2})&=&\sum\limits_{j=\rho,\omega}\,\sum\limits_{i=1,2}
 \frac{g_{\eta j \gamma}(M_\eta^2,M_\eta^2(1-2 x_i),0)g_{\pi j \gamma}(M_\pi^2,M_\eta^2(1-2 x_i),0)}
{G_{2}^{-1}-\Pi_{VV}^{uu}(M_\eta^2(1-2 x_i))},\\
 A(x_{1},x_{2})&=&\sum\limits_{j=\rho,\omega}\, \sum\limits_{i=1,2}
 \frac{g_{\eta j \gamma}(M_\eta^2,M_\eta^2(1-2 x_i),0)g_{\pi j \gamma}(M_\pi^2,M_\eta^2(1-2 x_i),0)M_\eta^2(1-x_i)}
{G_{2}^{-1}-\Pi_{VV}^{uu}(M_\eta^2(1-2 x_i))}.\nonumber
\end{eqnarray}

The box diagram is of a more complicated structure. It consists of three types of boxes (plus
three crossed) and contains the diagrams with pseudoscalar and axial-vector components of the
$\pi$ and $\eta$ mesons
\begin{eqnarray}
  T_{\mu\nu}=-i e^{2} \int \frac{d^4_\Lambda k}{(2\pi)^4}
  \mathrm {Tr}_D \biggl(\biggr. &&V_{\pi} S(k) V_{\eta}     S(k+p-q_1-q_{2}) \gamma_{\nu} S(k+p-q_1)   \gamma_{\mu} S(k+p)+\nonumber\\
&&+V_{\pi} S(k) V_{\eta}     S(k+q_2)         \gamma_{\nu} S(k+p-q_1)   \gamma_{\mu} S(k+p)\\
&&+V_{\pi} S(k) \gamma_{\nu} S(k+q_2)         \gamma_{\mu} S(k+q_1+q_2) V_{\eta}     S(k+p)
  +\left\{q_1 \leftrightarrow q_2 ,\mu \leftrightarrow \nu \right\}\biggl.\biggr)\nonumber
\end{eqnarray}
We calculate these diagrams numerically. In order to check the integration procedure, we
calculate all coefficients of different tensor structures and verify if they have gauge
invariant form (\ref{ampform}).

The obtained results for the decay width are given in the Table 1 for two sets of model
parameters. The main contribution comes from the box diagram. The contribution from vector
mesons has a constructive interference while the scalar $a_0$ contribution has a destructive
one. The results are in satisfactory agreement with Crystal Ball data $0.45\pm0.12$
\cite{Prakhov:2005vx} and the present value $0.57\pm0.21$ given in PDG\cite{PDBook}.

\begin{table}
\caption{$\eta\to\pi^0\gamma\gamma$ decay width.}
\begin{tabular}{|c|c|c|c|c|c|c|c|c|c|}
 \hline
 Contribution        &model 1& model 2  \\
 \hline
 vector mesons       & 0.17  & 0.20     \\
 scalar meson        & 0.03  & 0.12    \\
 vector+scalar mesons& 0.10  & 0.12    \\
 box                 & 0.28  & 0.35      \\
 box+vector          & 0.78  & 0.95    \\
 total               & 0.53  & 0.45 \\
 \hline
\end{tabular}
\end{table}

\begin{figure}
\resizebox{0.5\textwidth}{!}{\includegraphics{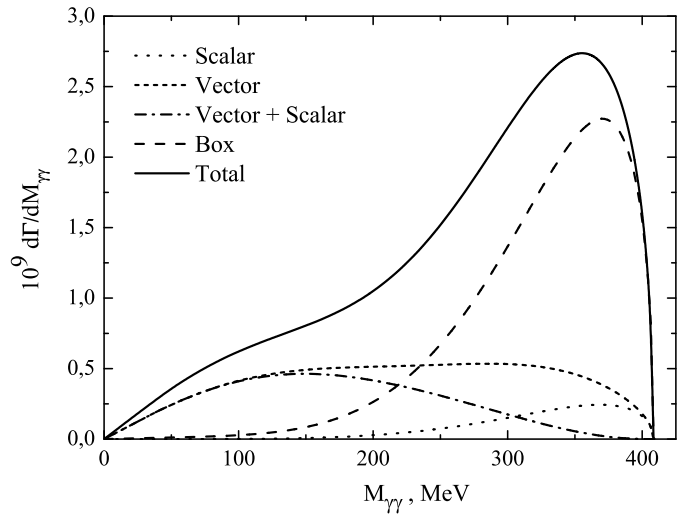}}
\caption{\label{fig:DifEtaPiGaGaS1} Invariant mass distribution of the two-photons of the
scalar meson contribution(dots), vector meson contributions(short dash), scalar + vector
mesons(dash-dot), quark box(long dash) and total(continuous line) for the set I.}
\end{figure}

\begin{figure}
\resizebox{0.5\textwidth}{!}{\includegraphics{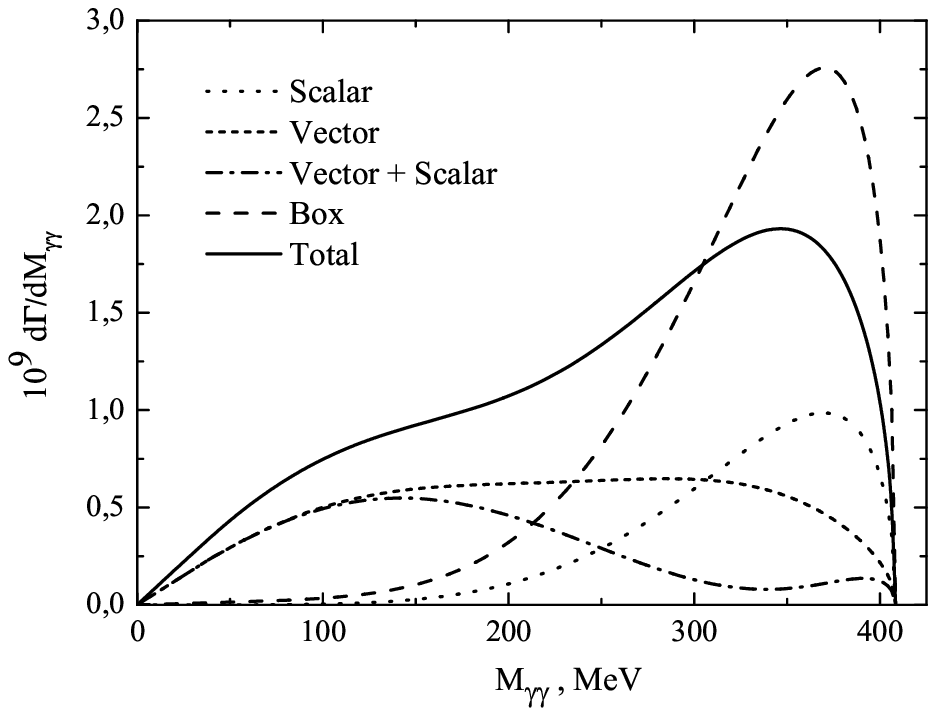}}
\caption{\label{fig:DifEtaPiGaGaS2} Invariant mass distribution of the two-photons of the
scalar meson contribution(dots), vector meson contributions(short dash), scalar + vector
mesons(dash-dot) , quark box(long dash) and total(continuous line) for the set II.}
\end{figure}

\begin{figure}
\resizebox{0.5\textwidth}{!}{\includegraphics{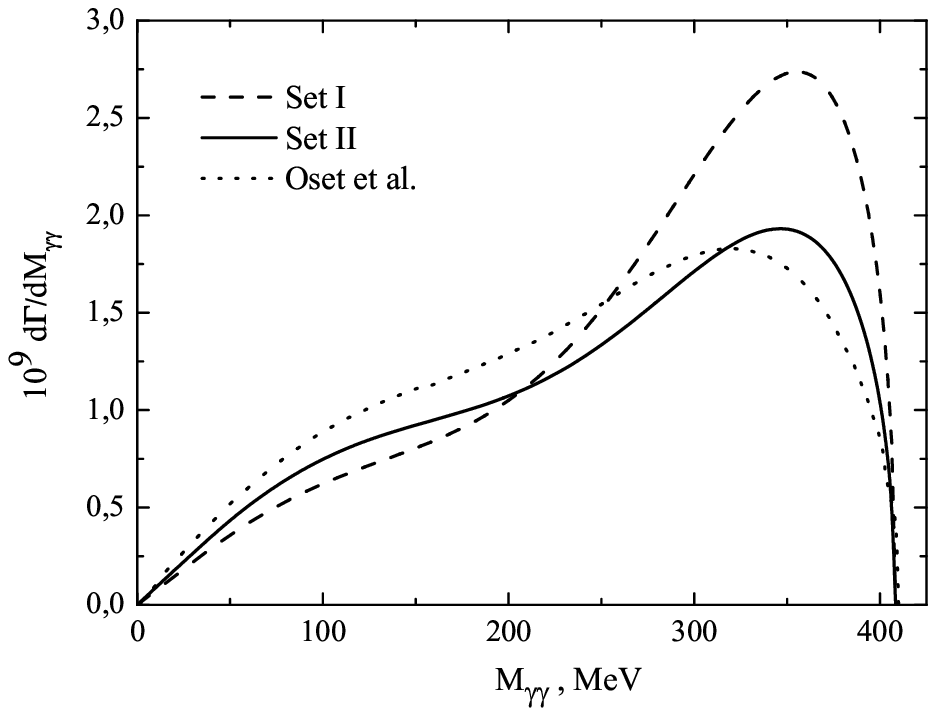}}
\caption{\label{fig:DifEtaPiGaGaS1S2Os} Invariant mass distribution of the two-photons of the
total contributions for the set I(dashes), set II(dots) together with the results of the
chiral unitary approach \cite{Oset:2002sh}.}
\end{figure}

It is also very instructive to consider the invariant mass distribution. In Figures
\ref{fig:DifEtaPiGaGaS1} and \ref{fig:DifEtaPiGaGaS2} the invariant mass distribution of the
two-photons is shown for the scalar meson contribution, vector mesons contribution, scalar +
vector mesons and total. In Figure \ref{fig:DifEtaPiGaGaS1S2Os}, the results of our
calculations of the invariant mass distribution are compared with the calculation in the
chiral unitary approach \cite{Oset:2002sh}.

\section{Conclusions}

Earlier calculations of the process $\eta \to \pi^0 \gamma\gamma$ in the NJL model do not
include the momentum dependence of quark loops and pseudoscalar--axial-vector transitions and
are in satisfactory agreement with the GAMS experiment.

Recently, the new experimental data on this decay have been obtained and the value of the
decay width is almost two times smaller. A number of theoretical estimates is also obtained,
and it seems that the momentum dependence of amplitudes is important for a correct description
of this process ( ``all-order'' estimate in ChPT).

In the present work, the contributions from quark box, scalar and vector pole diagrams are
considered with the full momentum dependence. The pseudoscalar--axial-vector transitions are
also taken into account.

The obtained result is consistent with recent experiments, theoretical estimates of ChPT
\cite{Ametller:1991dp,Bellucci:1995ay} and the chiral unitary approach\cite{Oset:2002sh}.

In future, we plan to consider the polarizability of pions and also decays of vector mesons
$\rho(\omega)\to\eta(\pi)\pi\gamma$.
\begin{acknowledgments}
The authors thank I. V. Anikin, A. E. Dorokhov, A. A. Osipov and V. L. Yudichev for useful
discussions. The authors acknowledge the support of the Russian Foundation for Basic Research,
under contract 05-02-16699.
\end{acknowledgments}

\section*{Appendixes}
\subsection{Lagrangian}

Lagrangian (\ref{Ldet}) can be rewritten in the form (see
\cite{Klimt:1989pm,Klevansky:1992qe})
\begin{eqnarray}
&&\mathcal{L} =
 {\bar q}(i{\hat \partial} - m^0)q +
  {\frac{1}{2}} \sum_{i=1}^9
   [G_i^{(-)} ({\bar q}{\lambda^\prime}_i q)^2 +G_i^{(+)}({\bar q}i{\gamma}_5{\lambda^\prime}_i q)^2] +
     \nonumber \\
  &&\qquad+ G^{(-)}_{us}({\bar q}           {\lambda}_u q)({\bar q}            {\lambda}_s q)
          + G^{(+)}_{us}({\bar q}i{\gamma}_5{\lambda}_u q)({\bar q}i {\gamma}_5{\lambda}_s q)
  + \frac{G_V}{2}\sum_{i=0}^8 [({\bar q}{\gamma}_\mu {         \lambda}_i q)^2
                              +({\bar q}{\gamma}_5{\gamma}_\mu{\lambda}_i q)^2],
\label{LGus}
\end{eqnarray}
where
\begin{eqnarray}
&&{\lambda^\prime}_i={\lambda}_i ~~~ (i=1,...,7),~~~\lambda^\prime_8 = \lambda_u =
({\sqrt 2}
\lambda_0 + \lambda_8)/{\sqrt 3},\nonumber\\
&&\lambda^\prime_9 = \lambda_s = (-\lambda_0 + {\sqrt 2}\lambda_8)/{\sqrt 3}, \label{DefG}\\
&&G_1^{(\pm)}=G_2^{(\pm)}=G_3^{(\pm)}= G \pm 4Km_sI_1 (m_s), \nonumber \\
&&G_4^{(\pm)}=G_5^{(\pm)}=G_6^{(\pm)}=G_7^{(\pm)}= G \pm 4Km_uI_1 (m_u),
\nonumber \\
&&G_u^{(\pm)}= G \mp 4Km_sI_1(m_s), ~~~ G_s^{(\pm)}= G, ~~~ G_{us}^{(\pm)}= \pm 4{\sqrt
2}Km_uI_1 (m_u).\nonumber
\end{eqnarray}

\subsection{Polarization loops}
Polarization loops in different channels after the PV regularization
\begin{eqnarray}
e^{-izm_im_j} \to R_{ij}(z)=e^{-izm_im_j}\left[1-(1+iz\Lambda^2)e^{-iz\Lambda^2}\right]
\end{eqnarray}
take the form (see \cite{Bernard:1995hm} for the expressions for the polarization loops with
equal indices)
\begin{eqnarray}
 \Pi_{PP}^{ij}(p^2) &=&\frac{N_c}{4\pi^2}\int_{-1}^1dy\int_0^{\infty}\frac{dz}{z}R_{ij}(z)e^{izA}
                \left[-\frac{i}{z}+\frac{1}{2}p^2(1-y^2)-\frac{1}{2}\left[(m_i-m_j)^2-y(m_i^2-m_j^2)\right]\right] ,\nonumber\\
 \Pi_{SS}^{ij}(p^2) &=&\Pi_{PP}^{ij}(p^2)-2m_im_j\frac{N_c}{4\pi^2}\int_{-1}^1dy\int_0^{\infty}\frac{dz}{z}R_{ij}(z)e^{izA} ,\nonumber\\
 \Pi_{VV}^{ij,{\mu\nu}}(p^2) &=&\left(g^{\mu\nu}-\frac{p^{\mu}p^{\nu}}{p^2}\right) \Pi_{VV}^{ij}(p^2) + \frac{p^{\mu}p^{\nu}}{p^2} \Pi_{VV}^{ij,L}(p^2),\nonumber\\
 \Pi_{VV}^{ij,L}(p^2) &=&\frac{N_c}{8\pi^2}\int_{-1}^1dy\int_0^{\infty}\frac{dz}{z}R_{ij}(z)e^{izA}
                \left[(m_i-m_j)^2-y(m_i^2-m_j^2)\right]\nonumber,\\
 \Pi_{VV}^{ij}(p^2) &=& \Pi_{VV}^{ij,L}(p^2)- p^2 \frac{N_c}{8\pi^2}\int_{-1}^1dy(1-y^2)\int_0^{\infty}\frac{dz}{z}R_{ij}(z)e^{izA}
                \nonumber,\\
 \Pi_{AA}^{ij,{\mu\nu}}(p^2) &=&\left(g^{\mu\nu}-\frac{p^{\mu}p^{\nu}}{p^2}\right) \Pi_{AA}^{ij,T}(p^2) + \frac{p^{\mu}p^{\nu}}{p^2} \Pi_{AA}^{ij}(p^2),\nonumber\\
 \Pi_{AA}^{ij,T}(p^2) &=&\Pi_{VV}^{ij}(p^2)+2m_im_j\frac{N_c}{4\pi^2}\int_{-1}^1dy\int_0^{\infty}\frac{dz}{z}R_{ij}(z)e^{izA},\\
 \Pi_{AA}^{ij}(p^2) &=&\Pi_{VV}^{ij,L}(p^2)+2m_im_j\frac{N_c}{4\pi^2}\int_{-1}^1dy\int_0^{\infty}\frac{dz}{z}R_{ij}(z)e^{izA}\nonumber,\\
 \Pi_{PA}^{ij,\mu}(p^2) &=& \frac{p^{\mu}}{\sqrt{p^2}} \Pi_{PA}^{ij}(p^2) = p^{\mu} i(m_i+m_j) \frac{N_c}{8\pi^2}\int_{-1}^1dy\int_0^{\infty}\frac{dz}{z}R_{ij}(z)e^{izA}\nonumber,\\
 \Pi_{AP}^{ij,\mu}(p^2) &=& \frac{p^{\mu}}{\sqrt{p^2}} \Pi_{AP}^{ij}(p^2) =-p^{\mu} i(m_i+m_j) \frac{N_c}{8\pi^2}\int_{-1}^1dy\int_0^{\infty}\frac{dz}{z}R_{ij}(z)e^{izA}\nonumber,\\
                &&A=\frac{p^2}{4}(1-y^2)-\frac{1}{2}\left[(m_i-m_j)^2-y(m_i^2-m_j^2)\right]\nonumber.
\end{eqnarray}

\subsection{Vertex functions}

The most simple form have the vertex functions for the vector $\rho$ and the isovector scalar
meson $a_0$, namely \footnote{We suppress flavor indices.}:
\begin{eqnarray}
V_{a_0}=g_{a_0} \mathbf{I} a_0, \quad V_{\rho}= g_{\rho} \gamma_\mu \rho^\mu.
\end{eqnarray}
The matrices $\mathbf{G}$ and $\mathbf{\Pi}$ for $a_0$ and $\rho$ mesons have the form
\begin{eqnarray}
\mathbf{G}_{a_0} &=& G_1^{(-)},\, \mathbf{\Pi}_{a_0}(p^2)  = \Pi_{SS}^{uu}(p^2), \\
\mathbf{G}_{\rho}&=&G_2  \quad     , \mathbf{\Pi}_{\rho}(p^2) = \Pi_{VV}^{uu}(p^2)\nonumber
\end{eqnarray}

For the pion and kaon,  additional axial-vector components appear in the vertex function due
to pseudoscalar--axial-vector mixing
\begin{eqnarray}
 V_{\pi}=g_{\pi} i \gamma_5 (1+\Delta_\pi \hat{p}) \pi,\,\quad
 V_{K}  =g_{K} i\gamma_5(1+\Delta_K \hat{p}) K
\end{eqnarray}
Here $\mathbf{G}$ and $\mathbf{\Pi}$ are
\begin{eqnarray}
\mathbf{G}_{\pi} =
\begin{pmatrix} G_1^{(+)}&0\\
                0        &G_2\end{pmatrix}
, \mathbf{\Pi}_{\pi}(p^2) =
\begin{pmatrix}
\Pi^{uu}_{PP}(p^2) & \Pi^{uu}_{PA}(p^2)\\
\Pi^{uu}_{AP}(p^2) & \Pi^{uu}_{AA}(p^2)
\end{pmatrix},\\
\mathbf{G}_{K} =
\begin{pmatrix} G_4^{(+)}&0\\
                0        &G_2\end{pmatrix}
, \mathbf{\Pi}_{K}(p^2) =
\begin{pmatrix}
\Pi^{us}_{PP}(p^2) & \Pi^{us}_{PA}(p^2)\\
\Pi^{us}_{AP}(p^2) & \Pi^{us}_{AA}(p^2)
\end{pmatrix}.\nonumber
\end{eqnarray}

Therefore, the vertex function of the $\eta$ meson have four components: strange and
non-strange pseudoscalar and axial-vector
\begin{eqnarray}
V_{\eta}&=&
  g_{\eta_u} i \gamma_5 (1+\Delta_{\eta_u} \hat{p}) \eta_u
 +g_{\eta_s} i \gamma_5 (1+\Delta_{\eta_s} \hat{p}) \eta_s =\\
 &=& g_{\eta} i \gamma_5( \cos\Theta_\eta\eta_u-\sin\Theta_\eta\eta_s +\Delta_{\eta} \hat{p}( \cos\widetilde{\Theta}_\eta \eta_u
 -\sin\widetilde{\Theta}_\eta\eta_s)),\nonumber
\end{eqnarray}
where $\Theta_\eta$ and $\widetilde{\Theta}_\eta$ are the mixing angles for pseudoscalar and
axial-vector components. The matrices $\mathbf{G}$ and $\mathbf{\Pi}(p^2)$ are four-by-four
matrices
\begin{eqnarray}
\mathbf{G} =
\begin{pmatrix} \mathbf{G}^{(+)}&0\\
                0        &\mathbf{G_2}\end{pmatrix}
, \mathbf{G}^{(+)} =
\begin{pmatrix} G_u^{(+)}&G_{us}^{(+)}\\
                G_{us}^{(+)}&G_s^{(+)}\end{pmatrix},\mathbf{G_2}=\mathrm{diag}\{G_2,G_2\}\\
\mathbf{\Pi}(p^2) =
\begin{pmatrix}
\mathbf{\Pi}_{PP}(p^2) & \mathbf{\Pi}_{PA}(p^2)\\
\mathbf{\Pi}_{AP}(p^2) & \mathbf{\Pi}_{AA}(p^2)
\end{pmatrix}
, \mathbf{\Pi}_{ij}(p^2) = \mathrm{diag}\{\Pi^{uu}_{ij(p^2)},\Pi^{ss}_{ij}(p^2)\},
i,j=P,A\nonumber
\end{eqnarray}

\subsection{Amplitudes $\eta\to\gamma\gamma$, $\rho\to\eta(\pi)\gamma$}

The amplitude for the two-photon decay width of the pseudoscalar meson has the form
\begin{eqnarray}
A(P\to\gamma\gamma) \ = \ e^2\; g_{P\gamma\gamma}(M_P^2,q_1^2,q_2^2)\
\epsilon_{\mu\nu\alpha\beta} \ \epsilon_1^{\mu} \epsilon_2^{\nu} \;q_1^\alpha
q_2^\beta\ ,
\end{eqnarray}
where $q_1$, $q_2$ are the momentum of photons and $\epsilon^{1}_\mu$, $\epsilon^{2}_\nu$ are
the polarization vectors of the photons,
\begin{eqnarray}
g_{\pi\gamma\gamma}(M_\pi^2,q_1^2,q_2^2) & = & I_u(M_\pi^2,q_1^2,q_2^2)g_{\pi},  \\
g_{\eta\gamma\gamma}(M_\eta^2,q_1^2,q_2^2) & = &\frac{5}{3}
I_u(M_\eta^2,q_1^2,q_2^2)g_{\eta_u}-\frac{\sqrt 2}{3} I_s(M_\eta^2,q_1^2,q_2^2)g_{\eta_s}
.\nonumber
\end{eqnarray}
The loop integrals $I_j(M_P^2)$ are given by
\begin{eqnarray}
I_j(M_P^2,q_1^2,q_2^2) & = &
 \frac{1}{2 \pi^2}  \int \limits_0^1 dx_1 \int \limits_0^{1-x_1} dx_2
 \frac{m_j}{m_j^2-x_1(1-x_1-x_2)q_1^2-x_2(1-x_1-x_2) q_2^2-x_1x_2 M_P^2}.
\end{eqnarray}
The amplitudes for the processes $\rho(\omega)\to\eta(\pi)\gamma$ have the form
\begin{eqnarray}
A(P V \gamma) \ = \ g_\rho e\; g_{P\rho\gamma}(M_P^2,q_1^2,q_2^2)\
\epsilon_{\mu\nu\alpha\beta} \ \epsilon_1^{\mu} \epsilon_2^{\nu} \;q_1^\alpha q_2^\beta\ ,
\end{eqnarray}
here $q_1$ and $\epsilon^{1}_\mu$ are the momentum and the polarization vector of
$\rho(\omega)$ meson.
\begin{eqnarray}
g_{\pi \rho\gamma}(M_\pi^2,q_1^2,q_2^2)  & = & I_u(M_\pi^2,q_1^2,q_2^2)g_{\pi},\quad
g_{\eta\rho\gamma}(M_\eta^2,q_1^2,q_2^2)  = 3I_u(M_\eta^2,q_1^2,q_2^2)g_{\eta_u}, \nonumber\\
g_{\pi \omega\gamma}(M_\pi^2,q_1^2,q_2^2)  & = &3I_u(M_\pi^2,q_1^2,q_2^2)g_{\pi} ,\quad
g_{\eta\omega\gamma}(M_\eta^2,q_1^2,q_2^2)  =  I_u(M_\eta^2,q_1^2,q_2^2)g_{\eta_u}.
\end{eqnarray}

%\nocite{*}

\end{document}